\documentclass{elsart}
\usepackage{graphics}
\usepackage{graphicx}
\usepackage{epsfig}
\usepackage{amssymb}
\usepackage{amsmath}

\newcommand{\dd}{\mbox{\textrm{d}}}

\begin{document}

\begin{frontmatter}

\title{Hard bremsstrahlung in the $\mathbf{pp\to pp\gamma}$ reaction}
\author[UU]{A.~Johansson} and
\author[CW_College]{C.~Wilkin\corauthref{cor1}}
\ead{cw@hep.ucl.ac.uk} \corauth[cor1]{Corresponding author.}
\address[UU]{Division of Nuclear and Particle Physics, Uppsala University,
Box 516, S-751~20 Uppsala, Sweden}
\address[CW_College]{Physics and Astronomy Department, UCL, London WC1E 6BT, UK}
\begin{abstract}
The $pp\to pp\gamma$ reaction has been measured at a beam energy of
310~MeV by detecting both final protons in the PROMICE-WASA facility
and identifying a missing-mass peak. For those events where the $pp$
excitation is less than 3~MeV, the final diproton is almost purely in
the $^{1\!}S_0$ state and, under these conditions, there is complete
coverage in the photon c.m.\ angle $\theta_{\gamma}$. The linear
behaviour observed in $\cos^2\theta_{\gamma}$ shows that there is
almost no influence of an $E2$ multipole at this energy, though the
$E1$ and $M2$ must be rather similar in size.
\end{abstract}

\end{frontmatter}

Hard bremsstrahlung produced through the two-body $pn\to d\gamma$
reaction has been studied for many years in either the direct or
inverse (photoabsorption) reaction. Much less is known about the
other \emph{elementary} case of $pp\to \{pp\}_{\!s}\gamma$, where the
$\{pp\}_{\!s}$ system is at very low excitation energy $E_{pp}$, such
that the final diproton is in the spin-singlet $S$--wave,
\emph{i.e.}\ in the $^{1\!}S_0$ state. The selection rules in the two
cases are very different; the production of an intermediate
$\Delta(1232)$ isobar is very important for the $np$ reaction whereas
for the $pp$ case the dominant $\Delta(1232)N$ intermediate
contribution is forbidden~\cite{Laget}. A comparison of $pn\to
d\gamma$ and $pp\to \{pp\}_{\!s}\gamma$ might therefore cast light on
these high energy bremsstrahlung processes.

Recent results on $pp\to \{pp\}_{\!s}\gamma$ have been published by
the COSY-ANKE collaboration at beam energies of $T_p=353$, 500, and
550~MeV~\cite{Komarov}. The events were selected by demanding that
the excitation energy of the two observed protons was less than
3~MeV. Although the coverage was limited to near-forward proton
angles, corresponding to $\cos\theta_{\gamma} > 0.95$, the data
seemed to indicate a forward dip, especially at the two higher
energies. The observed distribution in $E_{pp}$ was consistent with
that expected from a final state interaction (\emph{fsi}) in the
$^{1\!}S_0$ channel and the angular distribution in the $pp$ rest
frame was also isotropic, as required for an $S$ wave.

Most of the earlier experiments were carried out using pairs of
counters placed on either side of the beam line and, as a
consequence, they had little or no acceptance at small
$E_{pp}$~\cite{BREMS}. One exception was the COSY-TOF work at
300~MeV~\cite{TOF}, but comparatively few events were obtained at low
$E_{pp}$ and it was not possible to construct an angular distribution
for such a selection.

Attempts have been made to study the problem by looking at the
photoabsorption on $^{3}$He leading to two fast protons and a
``spectator'' neutron,
$^{3}$He$(\gamma,2p)n$~\cite{Emura,Niccolai,Audit}. Interpreting
these data in terms of photoabsorption on a bound diproton, it was
claimed that, at energies corresponding to $T_p \approx
400$--600~MeV, the reaction was dominated by an $E2$ transition.
Unfortunately, the fraction of events associated with quasifree
absorption was only about 5\% of the total~\cite{Emura} and so there
could be significant contamination arising from the much larger
absorption on quasi-deuteron pairs in the $^{3}$He nucleus. A further
cause for caution is that there is also a small fraction of $P$-wave
spin-triplet $pp$ pairs in $^{3}$He~\cite{Argonne}.

The possible observables in $pp\to \{pp\}_{\!s}\gamma$, and their
relation to the production amplitudes, have been very clearly spelled
out in Ref.~\cite{Jouni2}. Taking just the three lowest multipoles,
and neglecting possible contributions from high initial partial
waves, the differential cross section should be of the form:

\begin{equation}
\frac{\dd\sigma}{\dd\Omega} =
\frac{3}{8\pi}\left[|E1+M2|^2\sin^2\theta_{\gamma}
+2|E1-M2|^2\cos^2\theta_{\gamma}
+10|E2|^2\sin^2\theta_{\gamma}\cos^2\theta_{\gamma}\right]\!.
\label{fit1}
\end{equation}
where $|E1|^2$, $|M2|^2$, and $|E2|^2$ are proportional to the
contributions of the individual multipoles to the integrated cross
section.  It is clear from this that, in the absence of $E2$, the
differential cross section should be linear in
$\cos^2\theta_{\gamma}$. However, even in this case, one would
require photon polarisation measurements in order to isolate the
individual $|E1|^2$ and $|M2|^2$ terms. Furthermore, since
$|E1-M2|^2$ cannot be negative, the cross section should be
forward-peaked. Deviations from linearity would be a sign of the
influence of an $E2$ multipole.

Estimates within dynamical models~\cite{Jouni1,Kanzo} suggest that
the $E2$ term should be quite large in the $T_p > 200$~MeV region
and, if this is the case, the cross section could exhibit a forward
dip, as indicated by the higher energy COSY-ANKE data~\cite{Komarov}.
To investigate fully this one needs $pp\to \{pp\}_{\!s}\gamma$ data
over a much wider angular interval and this has proved possible to
obtain at 310~MeV by using the PROMICE-WASA facility~\cite{PW}
situated at the CELSIUS storage ring~\cite{CELSIUS} of the The
Svedberg Laboratory.

The $pp\to pp\gamma$ data reported here were collected simultaneously
with those for $pp\to pp\pi^0$~\cite{pizero}. The detector assembly
and the experimental techniques were therefore identical and the
analyses of the data differ only in minor details, so that we can
here be brief.

An internal gas-jet hydrogen target was used in conjunction with the
stored proton beam. By operating the electron cooler throughout the
experiment, the background was reduced and the counting rate
increased due to the larger beam-target overlap. The integrated
luminosity of $340\pm 35$~nb$^{-1}$ was found by comparing the
numbers of simultaneously measured elastically scattered proton
events with world cross section data, as described in
Ref.~\cite{pizero}.

Protons from $\pi^0$ production have a maximum laboratory polar angle
of around $18^{\circ}$. The exact value depends sensitively upon the
energy of the stored proton beam and its measurement determined that
$T_p=309.7\pm0.3$~MeV.

For the bremsstrahlung study reported here, only the protons in the
final state were used, even though the detector system was also
capable of measuring photons. After exiting the scattering chamber,
the protons passed through a forward window counter (FWC), a tracker,
a forward trigger hodoscope (FTH) and usually stopped in a forward
range hodoscope (FRH). The four-quadrant scintillator of the FWC
eliminated most of the beam halo background but using this meant that
the coincident protons had to appear in different quadrants in order
to be detected.

Angular information for the protons was extracted from the FTH and
most precisely from the tracker. The system covered a range in proton
polar angles $3^{\circ} <\theta_p< 22^{\circ}$. Due to a small
misalignment of the detector system with respect to the beam axis,
there was a small dependence on the azimuthal angle, which was taken
into account in the Monte Carlo analysis.

In order to ensure particle identification, it was further required
that both protons of an accepted event penetrated at least into the
second layer of the FTH, consisting of 24 spiral scintillator
segments. This meant that the minimum proton energy was 39~MeV. There
was no high energy limit since all the relevant protons stopped in
the second FRH scintillator or earlier.

As described in detail in Ref.~\cite{pizero}, the energy associated
with a proton track was obtained from a combination of the calculated
angle-dependent range up to the entrance of the stopping scintillator
and the measured light output of that detector. A few of the protons
stopped in the dead region between scintillators and in such cases
they were assumed to have an energy corresponding to the midpoint of
the dead layer.

A time signal was extracted for each proton from the first layer of
the FTH. After calibrating the individual detectors and correcting
for the times of flight, a time-difference $\Delta t$ spectrum was
obtained with a FWHM of 0.9~ns. Thus, by accepting only events with
$|\Delta t| < 1.8$~ns, the number of accidental coincidences was kept
to of the order of one percent.

The raw data were reduced to be stored in intermediate files and
these were already used to produce very preliminary
results~\cite{Jozef2}. Information was included on particle identity,
azimuthal and polar angles, energies, timing and energy loss in the
last detector of each track. About 60,000 events were seen in the
missing-mass peak attributed to the $pp\gamma$ final state.

\begin{figure}[htb]
\includegraphics[width=9cm]{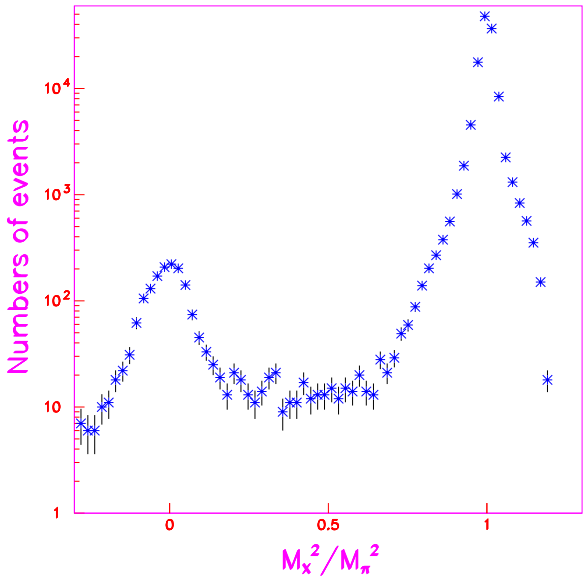}
\caption{Distribution in missing-mass squared of the $pp\to
\{pp\}_{\!s} X$ reaction for events with $E_{pp}<3$~MeV presented in
units of the neutral pion mass. Clear peaks are seen corresponding to
the $pp\to \{pp\}_{\!s} \pi^0$ and $pp\to \{pp\}_{\!s} \gamma$
reactions sitting on a slowly varying background.} \label{fig1}
\end{figure}

In a first step of the analysis, events were selected where the
excitation energy in the $pp$ rest system, $E_{pp}$, was less than
3~MeV. The square of the missing mass was evaluated for these events
and the resulting spectrum is shown in Fig.~\ref{fig1}. The data show
two clear peaks corresponding to the production of $\{pp\}_{\!s}
\pi^0$ and $\{pp\}_{\!s} \gamma$ final states. The $\gamma$-peak, of
width $\sigma(M_X^2)_{\gamma} \approx 0.06\,M_{\pi^0}^2$, contains in
total about 1450 events. This is sitting on a smoothly varying
background which is at about the 10\% level. To a good approximation,
this can be taken into account by keeping all events where
$|M_{X}^2/M_{\pi^0}^2| < 0.137$ and this criterium was applied in all
the angular bins. However, the background was larger for slow protons
and, rather than attempting to correct for this, data were only used
with $\cos\theta_{\gamma} < 0.8$. Due to the forward-backward
symmetry of the cross section, this did not result in any reduction
in the angular coverage.

In order to convert the observed number of events for given $E_{pp}$
and $\theta_{\gamma}$ values into cross sections, one needs to know
the detector acceptance as a function of these parameters. This was
achieved by Monte Carlo techniques, where the only deviation from
phase space was assumed to come from the $pp$ final-state-interaction
function discussed below. The detector system was described in great
geometric detail, with the simulated and experimental events being
required to pass the same tests. The acceptance was found to be quite
large, in most cases lying between 20\% and 30\%.

\begin{figure}[htb]
\includegraphics[width=9cm]{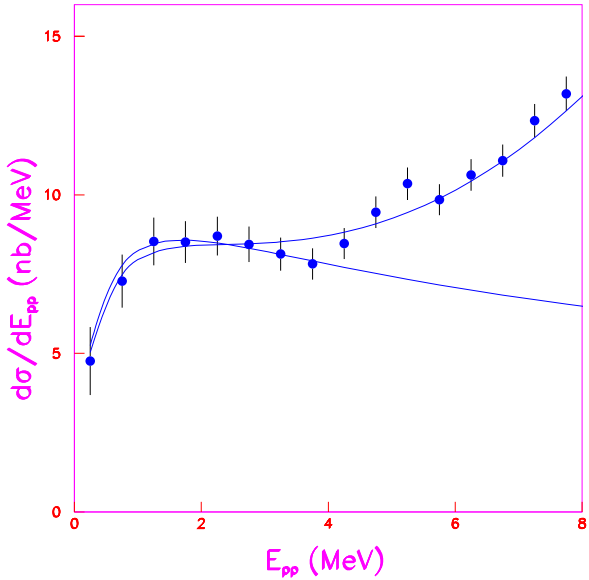}
\caption{Differential cross section for all $pp\to \{pp\}_{\!s}
\gamma$ events in terms of the excitation energy in the diproton. The
simulation of the shape of the $S$-wave \emph{fsi} enhancement is
based on Eq.~(\ref{PWF_fit}). } \label{fig2}
\end{figure}

The $pp\to\{pp\}_{\!s}\gamma$ differential cross section is shown
summed over all angles in Fig.~\ref{fig2} as a function of the
diproton excitation energy. The shoulder at small $E_{pp}$ is a clear
enhancement compared to phase space, which varies like
$\sqrt{E_{pp}}$ in this region. This is caused by the $S$-wave final
state interaction.

In the early Uppsala work on $\pi^0$ production~\cite{pizero1}, the
effect was evaluated in terms of the square of the $^{1\!}S_0$ $pp$
wave function at its peak ($r=1$~fm), divided by the corresponding
plane wave. In the case of the Paris wave function~\cite{Paris} the
enhancement factor may be parameterised for low $E_{pp}$ in the form
\begin{equation}
F_{\it fsi}(q)
=\frac{1+2.80q^2+20.28q^4-15.94q^6}{(1+51.02q^2)(1+14.36q^2)} \times
\frac{m\pi\alpha/q}{\exp(m\pi\alpha/q)-1}\,, \label{PWF_fit}
\end{equation}
where $m$ is the proton mass, $\alpha$ the fine structure constant,
and the $pp$ relative momentum $q=\sqrt{mE_{pp}}$ is measured in
fm$^{-1}$. It should be noted that no hard evidence is to be found
for the $pp$ \emph{fsi} in the COSY-TOF data~\cite{TOF}.

The enhancement factor of Eq.~(\ref{PWF_fit}) describes well the
shape of the differential cross section of Fig.~\ref{fig2} up to
about 4~MeV, when it seems that $P$ and higher waves start to become
more important. There is therefore likely to be very little
contamination to the $S$-wave if we only retain data for
$E_{pp}<3$~MeV. The direction of the $pp$ relative momentum vector is
hard to determine with precision for such small $E_{pp}$ but, just as
for the COSY-ANKE experiment~\cite{Komarov}, the distribution in this
vector is consistent with isotropy.

\begin{figure}[htb]
\includegraphics[width=9cm]{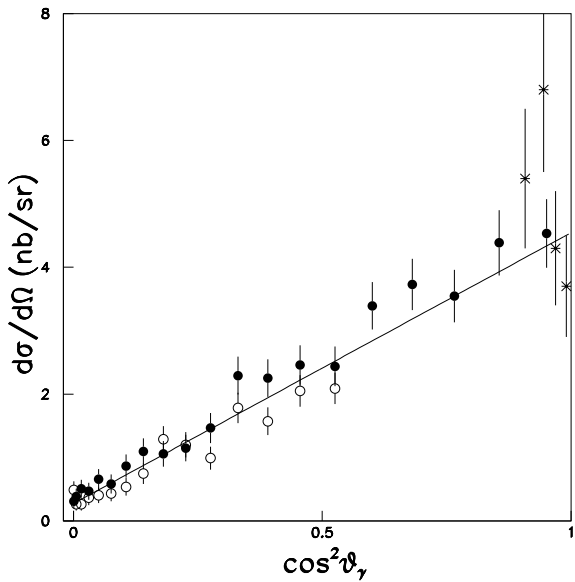}
\caption{Differential cross section for the $pp\to
\{pp\}_{\!s}\gamma$ reaction for $E_{pp}<3$~MeV as a function of
$\cos^2\theta_{\gamma}$. The present data at 310~MeV are shown by
closed circles for the backward photon hemisphere and open ones for
the forward. The COSY-ANKE results at 353~MeV~\cite{Komarov} are
denoted by crosses. The line represents a linear fit to both sets of
CELSIUS points.} \label{fig3}
\end{figure}

The angular distributions in both hemispheres are shown in
Fig.~\ref{fig3} in terms of $\cos^2\theta_{\gamma}$. These two data
sets are fairly consistent within the error bars, which reflects the
good description of the apparatus during the data analysis. Also
shown on this figure are the four points from the COSY-ANKE
collaboration~\cite{Komarov}. Although these were obtained at the
slightly higher energy of 353~MeV, they fall very close to our
results.

Fitting the CELSIUS angular distribution with the multipoles present
in Eq.~(\ref{fit1}) leads to
\begin{eqnarray}
\nonumber |E1+M2|^2 &=& \phantom{1}2.3\pm 0.5 \pm 0.3~\textrm{nb},\\
\nonumber |E1-M2|^2 &=& 18.9\pm 0.9 \pm0.8~\textrm{nb},\\
             |E2|^2 &<& 0.2~\textrm{nb},
\label{result}
\end{eqnarray}
where the first error is systematic, reflecting the slight
differences in results in the forward and backward hemispheres
apparent in Fig.~\ref{fig3}, and the second is statistical. There are
in addition overall systematic uncertainties of about 15\% that arise
principally from the luminosity determination, acceptance evaluation,
and background subtraction. The statistically best fit is obtained
with the negative value of $|E2|^2=-1.0\pm0.6$~nb and so only an
upper limit is quoted in Eq.~(\ref{result}) at the one standard
deviation level.

Although the individual contributions of the $E1$ and $M2$ multipoles
cannot be extracted from the data, it is clear from these results
that at 310~MeV $|E1|^2$ and $|M2|^2$ must be rather similar in size,
as indicated by theoretical estimates~\cite{Jouni1}. If we define the
ratio $M2/E1 = -r\textrm{e}^{i\phi}$, the data require that the phase
$|\phi|<41^{\circ}$ and the magnitude $0.46<r<2.2$, though this full
range is only allowed if $\phi$ is very small. A more rigorous bound
might be established if the phase were constrained by using the
Watson theorem. It is important to stress that there is no sign at
all of any significant $E2$ signal that was also predicted to be very
large~\cite{Jouni1} and further theoretical work in this area would
be most welcome.

On the other hand, there is evidence from the forward dip that there
must be large contributions from higher multipoles at 500 and
550~MeV~\cite{Komarov} and these might reflect in some form the
influence of the $\Delta(1232)$ isobar. The situation could be
clarified through measurements of proton analysing powers and spin
correlations~\cite{Jouni2} and this might be possible at
COSY~\cite{Kulikov}.

In summary, we have measured the $pp\to \{pp\}_{\!s}\gamma$
differential cross section over the full angular range for low
excitation energies in the $pp$ final state. The behaviour in
$E_{pp}$ is consistent with the belief that below 3~MeV the two
protons are almost entirely in the $^{1\!}S_0$ state. The photon
angular distribution shows that the $E1$ and $M2$ multipoles are
comparable in size but that, contrary to theoretical expectation,
$E2$ is quite small. The acceptance of the PROMICE-WASA apparatus is
very good for small $E_{pp}$ but the proton angular limitation to
$22^{\circ}$ in the laboratory system is equivalent to a maximum
possible $E_{pp}$ of 42~MeV. An analysis up to this limit will be
reported on at a later stage.

\vspace{0.5cm} We are very grateful to the TSL/ISV personnel for
their participation during the course of this work. The PROMICE-WASA
collaboration is acknowledged for the essential contributions
described in Ref.~\cite{pizero}. We are particularly grateful to
Jozef Z{\l}oma\'{n}czuk, who constructed the important intermediate
data files. Discussions and correspondence with K.~Nakayama on the
theoretical implications of the $pp\to \{pp\}_{\!s}\gamma$ reaction
were much appreciated. He also found an error in Eq.~\eqref{fit1} in
the published version of this paper, though this did not affect the
conclusions in any major way.  This work was supported by the
European Community under the ``Structuring the European Research
Area'' Specific Programme Research Infrastructures Action (Hadron
Physics, contact number RII3-cT-204-506078), and by the Swedish
Research Council.

\end{document}